# Electron-Hole Separation Dynamics and Optoelectronic Properties of a PCE10:FOIC Blend

G. Ammirati[1,*], S. Turchini[1], F. Toschi[1], P. O'Keeffe[1,2], A. Paladini[1,2], G. Mattioli[2], P. Moras[3], P. M. Sheverdyaeva[3], V. Milotti[3], C. J. Brabec[4,5], M. Wagner[4,5], I. McCulloch[6,7], A. Di Carlo[1,8], and D. Catone[1].

[1]*CNR-Istituto di Struttura della Materia (CNR-ISM), EuroFEL Support Laboratory (EFSL), Via del Fosso del Cavaliere 100, 00133, Rome, Italy.*

[2]*CNR-Istituto di Struttura della Materia (CNR-ISM), 00015, Monterotondo Scalo, Italy.*

[3]*CNR-Istituto di Struttura della Materia (CNR-ISM), SS 14, Km 163.5, I-34149, Trieste, Italy.*

[4]*Forschungszentrum Jülich GmbH, Helmholtz Institute Erlangen-Nürnberg for Renewable Energy (HI ERN), Dept. of High Throughput Methods in Photovoltaics, Erlangen, Germany.*

[5]*Friedrich-Alexander-Universität Erlangen-Nürnberg, Materials for Electronics and Energy Technology (i-MEET), Erlangen, Germany*

[6]*Andlinger Center for Energy and the Environment, and Department of Electrical and Computer Engineering, Princeton University, Princeton, NJ, 08544, USA*

[7]*Department of Chemistry, Oxford University, Chemistry Research Laboratory, Oxford OX1 3TA, U.K.*

[8]*CHOSE, University of Rome "Tor Vergata", Rome, 00133 Italy.*

*Corresponding author: giuseppe.ammirati@cnr.it

# Abstract

Understanding charge separation dynamics in organic semiconductor blends is crucial for optimizing the performance of organic photovoltaic solar cells. In this study, we explored the optoelectronic properties and charge separation dynamics of a PCE10:FOIC blend, by combining steady-state and time-resolved spectroscopies with high-level DFT calculations. Femtosecond transient absorption spectroscopy revealed a significant reduction of the exciton-exciton annihilation recombination rate in the acceptor when incorporated into the blend, compared to its pristine form. This reduction was attributed to a decrease in exciton density within the acceptor, driven by an efficient hole-separation process that was characterized by following the temporal evolution of the transient signals associated with the excited states of the donor when the acceptor was selectively excited within the blend. The analysis of these dynamics enabled the estimation of the hole separation time constant from the acceptor to the donor, yielding a time constant of (1.3 ± 0.3) ps. Additionally, this study allowed the quantification of exciton diffusion and revealed a charge separation efficiency of approximately 60%, providing valuable insights for the design of next-generation organic photovoltaic materials with enhanced charge separation and improved device efficiency.

# 1. Introduction

Organic photovoltaic (OPV) solar cells have emerged as a promising avenue for renewable energy production, offering a cost-effective alternative to traditional silicon-based solar cells. Their qualities, including flexibility, lightness, semi-transparency, and ease of processing, make them attractive for various applications such as outdoor and building integrated photovoltaics, and wearable electronic devices[1–4]. To further improve the performance of such cells, a comprehensive understanding of the intricate mechanisms governing the photovoltaic process is highly desirable. In bulk heterojunction (BHJ) architectures, excitons typically dissociate at the donor-acceptor (D-A) interface through charge separation (CS) processes, leading to the formation of free charges that are subsequently extracted as photocurrent.[5–7] The energy offset between donor and acceptor affects the open-circuit voltage; a small energy level offset is preferred to maximize the open circuit voltage, but this, on the other hand, can induce slower CS rates and increased exciton recombination, potentially reducing current generation.[8,9] Thus, achieving an optimal balance in the energy offset is crucial for enhancing the efficiencies in OPV systems. Fullerene-based literature has suggested that a minimum level offset of at least 0.3 eV is necessary for efficient CS. [8,10–12] [8,13] The introduction of non-fullerene acceptors (NFAs) into BHJ structures has recently led to a significant performance boost,[14–16] achieving record efficiencies close to 20%[17]. This success is attributed to their high absorption in the near-infrared region, even with minimal energy offset. Recent findings have shown that when the highest occupied molecular orbital (HOMO) offset drops below 0.2 eV, the CS rate decreases. This leads to a lower short-circuit current, which overcompensates for the open-circuit voltage gain.

Femtosecond Transient Absorption Spectroscopy (FTAS) is a powerful tool for investigating photophysical processes occurring in the active materials for OPV applications. This makes it particularly relevant for studying recombination dynamics and charge transfer at D-A bulk heterojunctions[12,18–24]. By offering a quantitative and time-resolved analysis of charge generation, transfer, and recombination dynamics, FTAS provides critical insights that drive advancements in the understanding and optimization of OPV materials, ultimately contributing to enhanced device performance. Remarkably, by enabling the direct observation of transient species such as singlet and triplet excitons, charge transfer and CS states, and free carriers, FTAS allows the identification and characterization of excitonic and charge-related processes, crucial for understanding the dissociation of excitons at the D-A interface and the generation of free carriers. By resolving the temporal evolution of these species, FTAS distinguishes between geminate recombination of tightly bound electron-hole pairs[19,25–27] and bimolecular recombination of independently generated charges[28]. Furthermore, its ability to monitor processes across a broad timescale, from femtoseconds to microseconds, creates a detailed analysis of both ultrafast charge transfer events[29–33] and longer-lived recombination pathways[34].

In this context, the ultrafast separation dynamics of both electrons and holes in BHJs were investigated with different configurations, demonstrating that the intrinsic hole separation remains on the picosecond time scale, even for a near-zero driving force[12,32]. However, the crucial connection between recombination losses and CS processes is missing. Whether recombination occurs geminately, where an electron and a hole recombine without ever leaving their point of origin, or through intermolecular interactions, it ultimately leads to a decrease in power conversion efficiency[35]. In the context of D-A systems, FTAS also serves as a diagnostic tool for assigning spectral features to specific components, allowing the evaluation of the component-dependent efficiency of CS and charge collection. This capability is essential for optimizing the design of bulk heterojunctions and improving the understanding of energy losses that limit device performance.

In this work, we employ a combination of advanced experimental techniques, assisted by atomistic simulations, to the study of the optoelectronic properties and CS dynamics of a D-A blend composed of PCE10 (also known as PTB7-Th) as D and FOIC as A. Such blend is characterized by a high transparency in the visible spectrum and by promising power conversion efficiencies, making it ideal for use as active material in semitransparent solar cells[18,36–38]. Through steady-state and time-resolved spectroscopic techniques, supported by high-level DFT calculations, we provide an in-depth analysis of absorption spectra, band diagrams, and excited-state properties of one-component thin films as well as blends, enabling us to explore the exciton-exciton annihilation dynamics and the hole separation mechanism between the A and D within the blend.

## 2. Methods

### 2.1. Materials

The samples investigated have a layered structure composed of glass/ITO/ZnO/PCE10:FOIC stacks. Indium tin oxide (ITO)-coated glass substrates with a sheet resistance of 15 Ω/□ were purchased from VisionTek. Zinc oxide (ZnO) nanoparticles (2.5 wt% in isopropanol) were purchased from Avantama. PCE10 (aka PTB7-Th) (Poly[4,8-bis(5-(2-ethylhexyl)thiophen-2-yl)benzo[1,2-b;4,5-b']dithiophene-2,6-diyl-alt-(4-(2-ethylhexyl)-3-fluorothieno[3,4-b]thiophene-)-2-carboxylate-2-6-diyl)]) was produced by KAUST. FOIC[39] was purchased from 1 material.

Glass/ITO substrates were cleaned in an ultrasonic bath. ZnO nanoparticles were coated with a doctor blade and annealed for 30 min at 200 °C in an ambient atmosphere, followed by the active layer. PCE10, FOIC, and PCE10:FOIC were diluted in chloroform (22.25 mg/ml for the single components and the Blend 1:1) and stirred at RT overnight before being coated with a doctor blade using the following processing parameters: 60 μL of solution (stirred at RT) were injected into a 400-μm gap between substrate and blade (both heated to 30 °C). Coating with 10 mm/s forms a wet film that dries very fast (2–3 s) due to the low boiling point of chloroform (61 °C). The thin films were subsequently annealed for 4 min at 140 °C under an inert atmosphere. The thickness of the films was estimated with a Brucker DecktatXT stylus profilometer to be 50 nm for FOIC, 200 nm for PCE10, and 105 nm for the Blend.

### 2.2. Photoelectron Spectroscopy

Photoelectron Spectroscopy (PES) measurements were performed at the VUV-Photoemission beamline (Elettra, Trieste) under ultra-high vacuum conditions at room temperature using a Scienta R-4000 electron analyzer. The photon energy and total energy resolution were set to 75 eV and 30 meV, respectively. The surface of the organic films was sputtered by Ar ions with an energy of 500 eV for 5 min to remove the surface contamination. The energy scale of the spectra was referenced to as the Fermi level of the Mo metallic sample holder in electrical contact with the studied films.

## 2.3. Transient Absorption Spectroscopy

The pump–probe experiments were performed by using a laser system consisting of a regenerative amplifier (Coherent Legend Elite Duo HE+), seeded by a Ti:Sa oscillator (Coherent Vitara-T), that generate pulses at 800 nm (1 kHz, 4 mJ) with 35 fs of duration. The pump pulse was produced by an Optical Parametric Amplifier (OPA) at selected photon energies. Conversely, the probe beam was generated by focusing 100 µJ of the regenerative amplifier output at 800 nm into a homemade OPA that generates an output beam at 1200 nm. This femtosecond radiation is focused on a sapphire crystal to generate a white light supercontinuum probe in the VIS-IR (500–1000 nm). The optical layout of the commercial transient absorption spectrometer (FemtoFrame II, IB Photonics) consisted of a split beam configuration in which 50% of the white light passes through the sample while the remainder was used as a reference to account for pulse-to-pulse fluctuations in the white light generation. The pump and the probe beams were focused on the sample with a diameter of 200 and 150 µm, respectively, and the delay time between the two was changed by modifying the optical path of the probe, resulting in an instrument response function of about 50 fs. All experiments were performed with linear polarization for both pump (vertical) and probe (horizontal) pulses. The pump scattering is removed by means of a linear polarizer located after the sample. More experimental details on the setup can be found elsewhere.[40,41]

## 2.3. DFT

The properties of FOIC, PCE10, and their blends were investigated using a multilevel computational protocol. A preliminary broad screening of molecular configurations has been performed for each system using a conformers-rotamers ensemble search tool (CREST[42]) based on the GFN-FF force field[43], which performs a complex combination of (meta)dynamics simulations and geometry optimizations, blended by genetic sorting and mixing of structures. All the systems are isolated but immersed in a low-polarity implicit dielectric environment to mimic the embedding into an organic film. A large ensemble of low-energy configurations (up to hundreds of systems) is then fully reoptimized and sorted in energy using a semiempirical tight-binding method rooted on the GFN2-xTB Hamiltonian, as implemented in the xTB suite of programs[44]. The lowest-energy structures were then investigated using (time-dependent) DFT-based simulations, carefully balanced between accuracy and feasibility as applied to large systems up to 2720 electrons. Geometry optimizations were performed in a plane-wave/pseudopotential framework using the Quantum ESPRESSO suite of programs[45]. All the systems are optimized in very large supercells in periodic boundary conditions using the C09[46] gradient corrected functional coupled with an ab initio VDW-DF [47,48] correction (C09+VDWDF). The C09+VDWDF method provides a reliable description of structures and interaction energies of noncovalent systems. The Kohn-Sham orbitals are expanded on a plane-wave basis set, with core electrons replaced by ultrasoft pseudopotentials[49]. Satisfactorily converged results have been obtained using 40 Ry (320 Ry) cutoffs on the plane waves (electronic densities). Accurate electronic and optical properties are then calculated on the obtained structures using a different DFT framework based on all-electron Gaussian-type basis sets as implemented in the ORCA suite of programs[50]. Single-point calculations were performed using the dispersion-corrected[51] B3LYP functional[52] and the def2-TZVPP all-electron basis set[53,54] The corresponding def2/J basis has also been used as an auxiliary basis set for Coulomb fitting in a resolution-of-identity/chain-of-spheres (RIJCOSX) framework. Absorption spectra were also calculated on C09+VDWDF geometries in the sTDDFT framework[55] using the same B3LYP functional and the same basis sets for individual components and for a FOIC/PCE10 diad in both trimer and tetramer cases. An infinite periodic PCE10 polymer chain has been used as a reference to ensure the convergence of the electronic level of isolated oligomers. This calculation has been performed in the plane-wave/pseudopotential framework discussed above using the same B3LYP functional. Optimized norm-conserving Vanderbilt pseudopotentials[56] have been used in this case to replace core electrons with 80/320 Ry cutoffs. For further details, see Section 1 of Supporting Information (SI).

# 3. Results and discussion

## 3.1. Steady-state optical and electronic properties

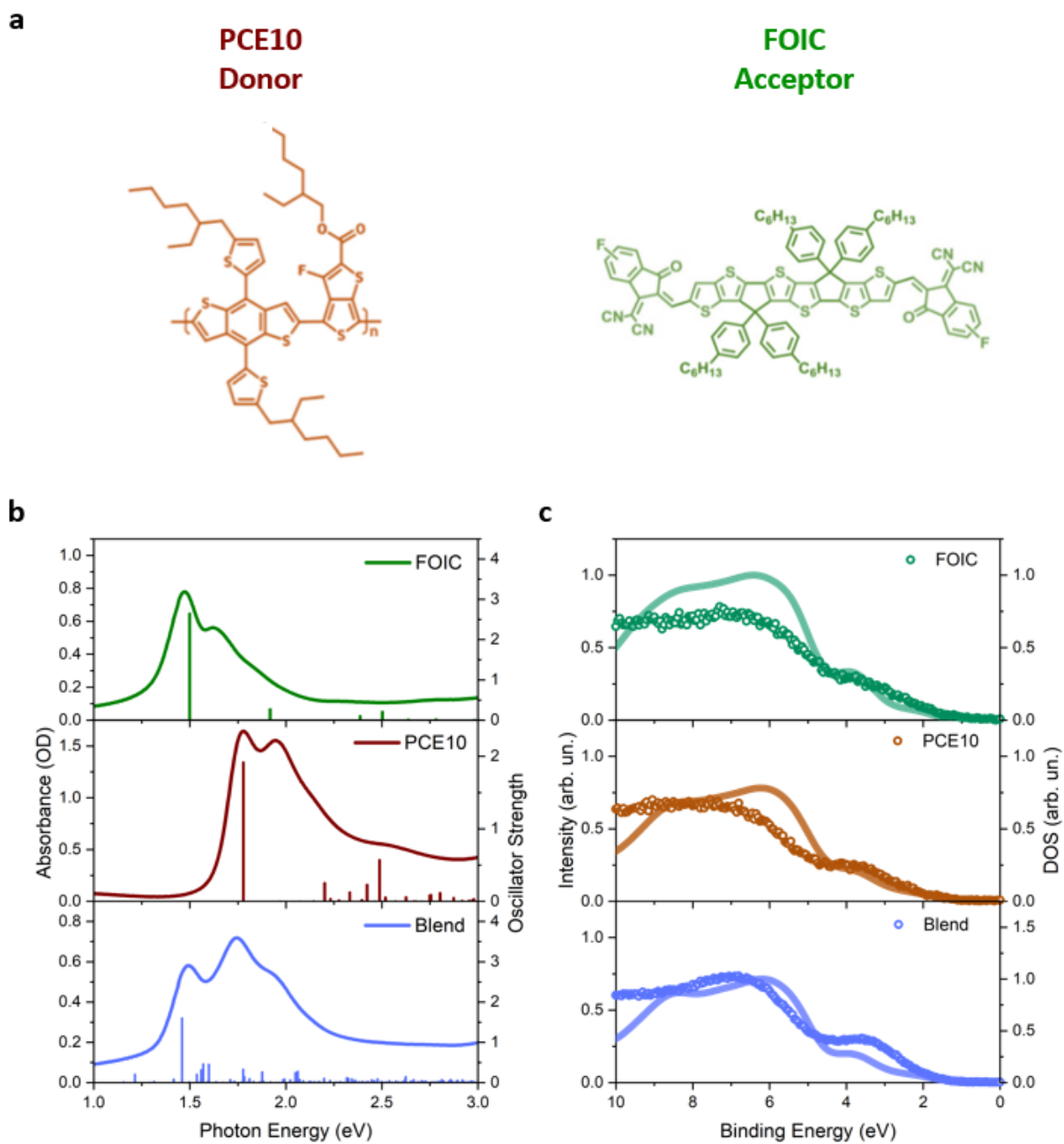

**Figure 1:** (a) Molecular structure of PCE10 (D) and FOIC (A); (b) experimental (line) and theoretical (bars) absorption spectra of FOIC (black), PCE10 (red), and Blend (blue) together with the spectra profiles of PCE10 (D) and FOIC (A). The theoretical results of PCE10 were rigidly shifted by +390 meV to align them with the experiment. (c) Experimental photoelectron spectra of the film of FOIC (green scatter), PCE10 (orange scatter), and Blend (blue scatter) acquired at the photon energy of 75 eV along with the corresponding theoretical DOS (lines).

The molecular structures of FOIC (A) and PCE10 (D) are shown in **Figure 1a**. The absorption spectra of FOIC (A), PCE10 (D), and PCE10:FOIC (Blend), acquired with a PerkinElmer Lambda19 spectrophotometer are shown in **Figure 1b** together with the optical transitions predicted by DFT and reported as bar diagrams (see **Section S1** of **SI** for further details). FOIC shows a broad absorption peak at 1.45 eV with additional structures at 1.7 and 1.8 eV. PCE10 exhibits a broad absorption feature with structures at 1.72, 2.0, 2.2, and 2.5 eV. The absorption spectrum of the Blend shows features that can be attributed to both A and D contributions. In particular, it is possible to assign the peak at 1.52 eV to a main A contribution, although slightly blue-shifted with respect to the bare FOIC, while the shoulders at approximately 1.7 and 2.0 eV resemble the D spectral features. Furthermore, the tails at higher energies in the experimental spectra are likely attributed to vibrational states of the molecular systems, an aspect not taken into account by the calculations. In the case of PCE10, the simulations give the first absorption peak at 1.39 eV for the tetramer (see **Table S1**), which is significantly redshifted compared to the experimental peak at 1.72 eV. Thus, the theoretical results obtained for PCE10 and shown in **Figure 1b** are rigidly shifted by +390 meV to align them with the experimental spectrum. This discrepancy likely arises from aggregation effects between polymer chains, not implemented in our simulations, which lead to an underestimation of the calculated absorption energy features. The optical band gaps ($E_g$) for the studied materials were estimated from the acquired absorption spectra, resulting in: 1.26 eV for FOIC; 1.58 eV for PCE10; 1.32 eV for Blend (see **Section S2** of SI for further details).

**Figure 1c** shows the PES spectra in the Valence Band (VB) region, obtained with a photon energy of 75 eV for FOIC, PCE10 and Blend (scattered point) together with the calculated DOS (solid lines), generated by the convolution of B3LYP Kohn-Sham eigenvalues with Gaussian functions ($\sigma$=32 meV) and shifted by +3.0 eV to match the experimental data. The theoretical predictions are in satisfactory agreement with the experimental data. The estimation of the binding energy onset was used to evaluate the VB energy with respect to the vacuum energy of the studied materials: 1.07 eV for FOIC; 0.97 eV for PCE10; 1.15 eV for Blend. The energy of the Conduction Band (CB) was then calculated by adding to the VB the $E_g$ obtained from the absorption spectra. In this way the band diagrams reported in **Figure 2** (for the details see **Section S2** of SI) were generated.

These results, which are in good agreement with the theoretical band diagram shown in **Figure S1**, allowed us to estimate the energy offset between the lowest unoccupied molecular orbital (LUMO) of D and the HOMO of A, which is an important parameter that strongly influences the optoelectronic properties of these materials and their performance when used for solar cell devices. Based on the experimental band diagram, the energy difference between the PCE10 VB maximum (-5.48 eV) and the FOIC CB minimum (-4.45 eV) yields an offset of 1.03 eV. Additionally, the HOMO offset and LUMO offset, determined by the energy difference between FOIC and PCE10, are 0.23 eV and 0.55 eV, respectively[57]. These results are consistent with other NFAs developed for photovoltaic applications, which have demonstrated good performances[58] with reduced losses that positively impact open-circuit voltage and fill factor.[59]

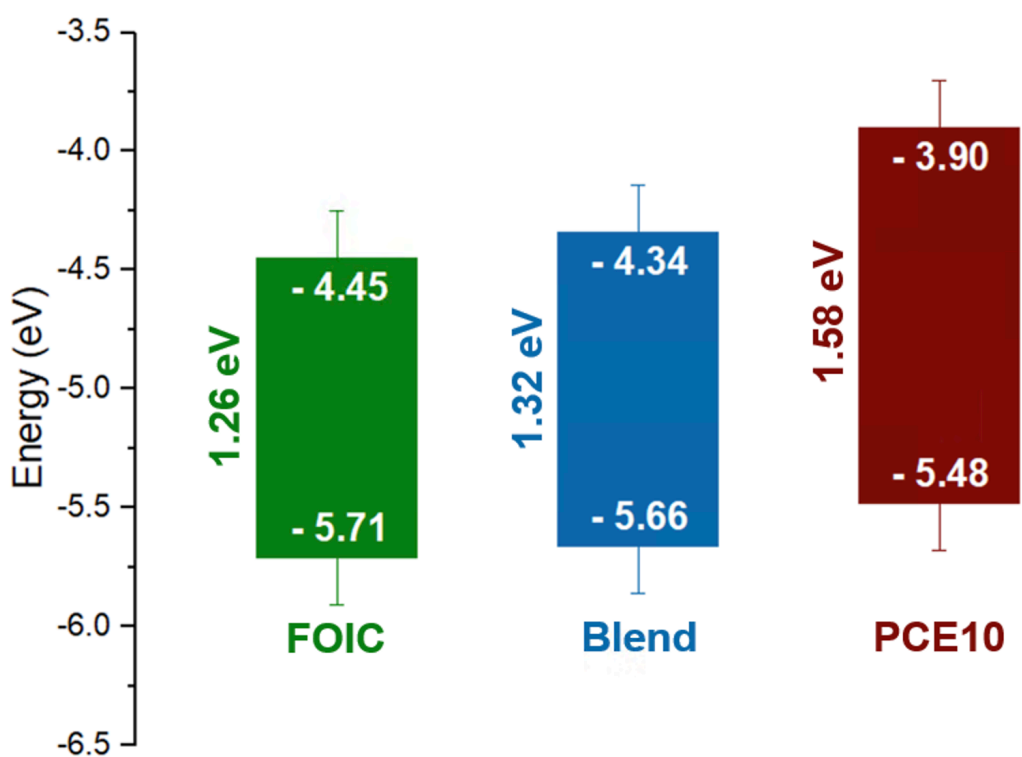

**Figure 2:** Experimental band diagram of FOIC, Blend, and PCE10. The energy of the valence band was evaluated by the onset of the PES spectra obtained at 75 eV, while the energy of the conduction band was then calculated by adding to the valence band the $E_g$ obtained from the absorption spectra. The energy values reported were estimated with respect to the vacuum energy.

## 3.2. Transient absorption spectroscopy and ultrafast dynamics in FOIC, PCE10, and Blend

On the basis of the ground-state electronic structure of FOIC, PCE10 and Blend reported above, it is possible to discuss the results obtained from the FTAS. **Figure 3a** shows the transient absorption (TA) spectra of FOIC and Blend obtained with a pump at 1.45 eV and a fluence of 220 µJ/cm$^2$, and the TA spectrum of PCE10 acquired with a pump at 1.75 eV. All the TA spectra show photobleaching (PB) signals (negative signals) that correspond to the peaks present in the absorption spectra. We observed a distinct behavior in the Blend when excited by a pump at 1.75 eV (see **Figure S4** in SI) and at 1.45 eV. When excited at 1.75 eV, the pump energy is sufficient to excite both the A and D components in the Blend, resulting in PB signals that resemble features in the absorption spectrum. In contrast, when excited at 1.45 eV, the pump energy can only excite the A component dispersed in the Blend (see spectra in **Figure 1a**), leading to TA spectra that closely resemble those of the pure FOIC. Additionally, the broad PB signal at probe energies around 1.9 eV, corresponding to the D contributions, shows a slower rise compared to other lower-energy PB signals. The nature and the dynamics of this PB feature will be discussed in detail later in the text.

In this context, we have studied the time dependence of the TA signals as a function of the excitation fluence, and consequently, as the exciton density increases. **Figure 3b** shows the decay dynamics of bleaching signals, labeled as PB1, for FOIC, PCE10 and Blend at selected excitation fluences for pump energies at 1.75 eV for PCE10 and 1.45 eV for FOIC and Blend (for the TA spectra obtained in other excitation conditions, see **Section S3** in the SI). The data show a decay trend in the first picoseconds, which becomes faster as the exciton density increases, suggesting that the dynamics is influenced by a scattering process. This trend was assigned to an exciton-exciton annihilation (EEA) process, i.e., a non-radiative many-body interaction that involves the energy transfer from one exciton to another[60]. In fact, this recombination process is described as a bimolecular process and was established as the dominant recombination mechanism within organic materials and in systems with reduced dimensionality[41,61–63] (see **Section S2** in the SI for further details on the data fitting procedure).

Although the exciton density used in this work is higher than that of OPV devices under typical operating conditions, where exciton splitting is the main process, estimating the EEA is of significant interest. This rate serves as a valuable tool for evaluating the excitonic diffusion coefficient ($D_{ex}$), a crucial parameter in the development of efficient organic photovoltaic devices. The relationship between EEA and D is as follows:[64]

$$D_{ex} = \frac{\gamma}{4\pi R} \qquad 1$$

where $\gamma$ is the EEA rate constant and $R$ is the exciton annihilation radius (approximated to 1 nm, as estimated in both experimental[65] and theoretical[66] studies).

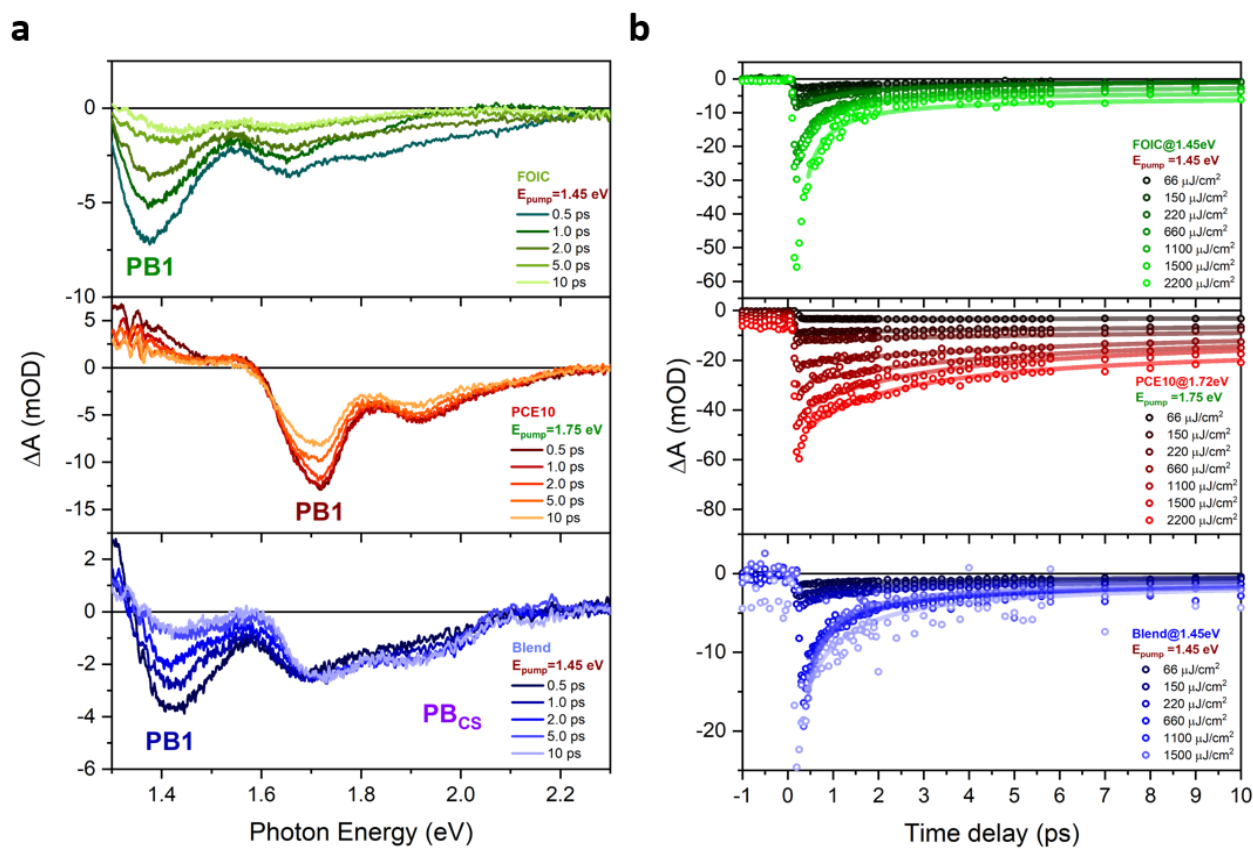

**Figure 3:** (a) TA spectra obtained at the pump fluence of 220 μJ/cm² for FOIC (green lines) at the pump photon energy of 1.45 eV, PCE10 (red lines) at the pump photon energy of 1.75 eV, and Blend (blue lines) at the pump photon energy of 1.45 eV. (b) Experimental (scatter) and fit (line) of the temporal dynamics of PB1 signals at selected fluences for FOIC (green), PCE10 (red), and Blend (blue) in the same experimental conditions. The non-zero signals at negative time delays are due to pump scattering.

**Figure 4** reports the EEA rates for FOIC (green), PCE10 (red), and Blend (blue), together with the corresponding linear fit used to estimate the γ values for the studied materials. The estimated EEA rates and diffusion constant for FOIC, PCE10, and Blend are reported in **Table 1**, and are comparable to those measured for other organic materials used in photovoltaics[41,67,68]. Here we highlight that, by exciting Blend with a pump at 1.45 eV, we are mainly observing the properties of the A (FOIC) molecule when dispersed in the Blend (see **Figure 5a**).

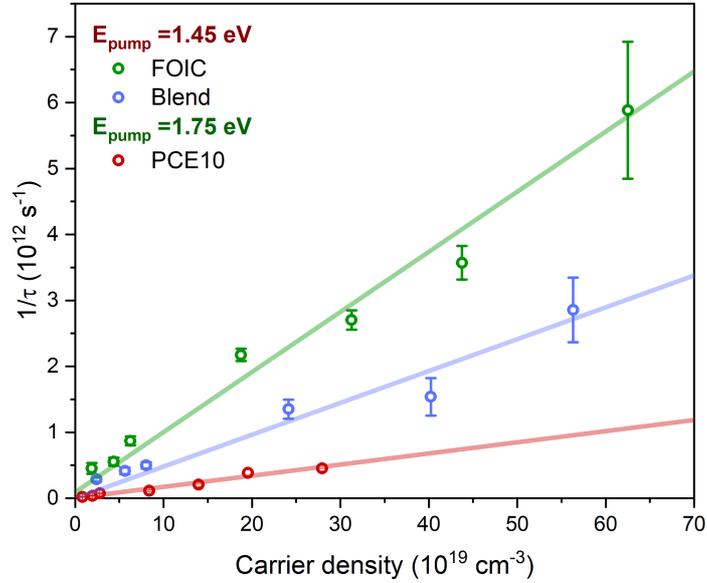

**Figure 4:** Experimental data (scatter) and fitting curve (line) of the EEA rate for FOIC (green), PCE10 (red), and Blend (blue). The linear fit was used to estimate the EEA rate constant γ for the studied materials.

|  | FOIC | PCE10 | A (FOIC) in Blend |
|---|---|---|---|
| γ ($10^{-9}$ cm$^3$ s$^{-1}$) | 9.1±0.4 | 1.7±0.2 | 4.8±0.4 |
| $D_{ex}$ ($10^{-3}$ cm$^2$ s$^{-1}$) | 7.2 | 1.4 | 3.8 |

**Table 1:** EEA rate constant γ and estimated diffusion coefficient D for FOIC, PCE10, and Blend.

The lower γ and the consequent lower $D_{ex}$ exhibited by PCE10 could be attributed to a lesser structural order compared to FOIC[67]. Conversely, the lower γ of the A (FOIC) within the Blend compared to pure FOIC indicates how the EEA recombination process is influenced by the interaction with the D (PCE10). Specifically, since EEA is a scattering process, the reduction in its rate constant suggests a decrease in the exciton density of FOIC within the Blend that may be partially due to CS, namely the hole displacement process from A to D. It is worth noting that the EEA rate was not evaluated for the D in the Blend, as it is impossible to excite PCE10 exclusively without also exciting FOIC.

## 3.3. Charge Separation Dynamics in FOIC:PCE10 blend

To further investigate electron-hole separation, we analyzed the broad PB signal, labeled as $PB_{CS}$, appearing in the 1.8-2.2 eV energy range a few picoseconds after the excitation of the Blend. This signal is almost entirely attributed to spectral features related to the PCE10 component (see **Figure 3a** and **Figure 5a**). This process effectively leads the PCE10 in a CS state, leading to a reduced population of PCE10 in its ground state after the FOIC photoexcitation (see the scheme in **Figure 5b**). Thus, following the temporal trend of the PB signal at about 1.9 eV, it was possible to estimate the time constant of the hole separation process that occurs in the Blend when only the A is excited, as reported in **Figure 5b**. The dynamics were fit by a rising exponential equation convoluted with a Gaussian function simulating the IRF and a step function (see **Section S3** in SI for further details), resulting in a time constant of (1.3±0.3) ps. This result is in excellent agreement with the hole separation time constant observed in similar blend systems, showing similar energy offsets[58]. Moreover, this PB signal shows a progressive broadening with the time delay, that can be ascribed to the incremental population of the quasi-free charge separated band[7].

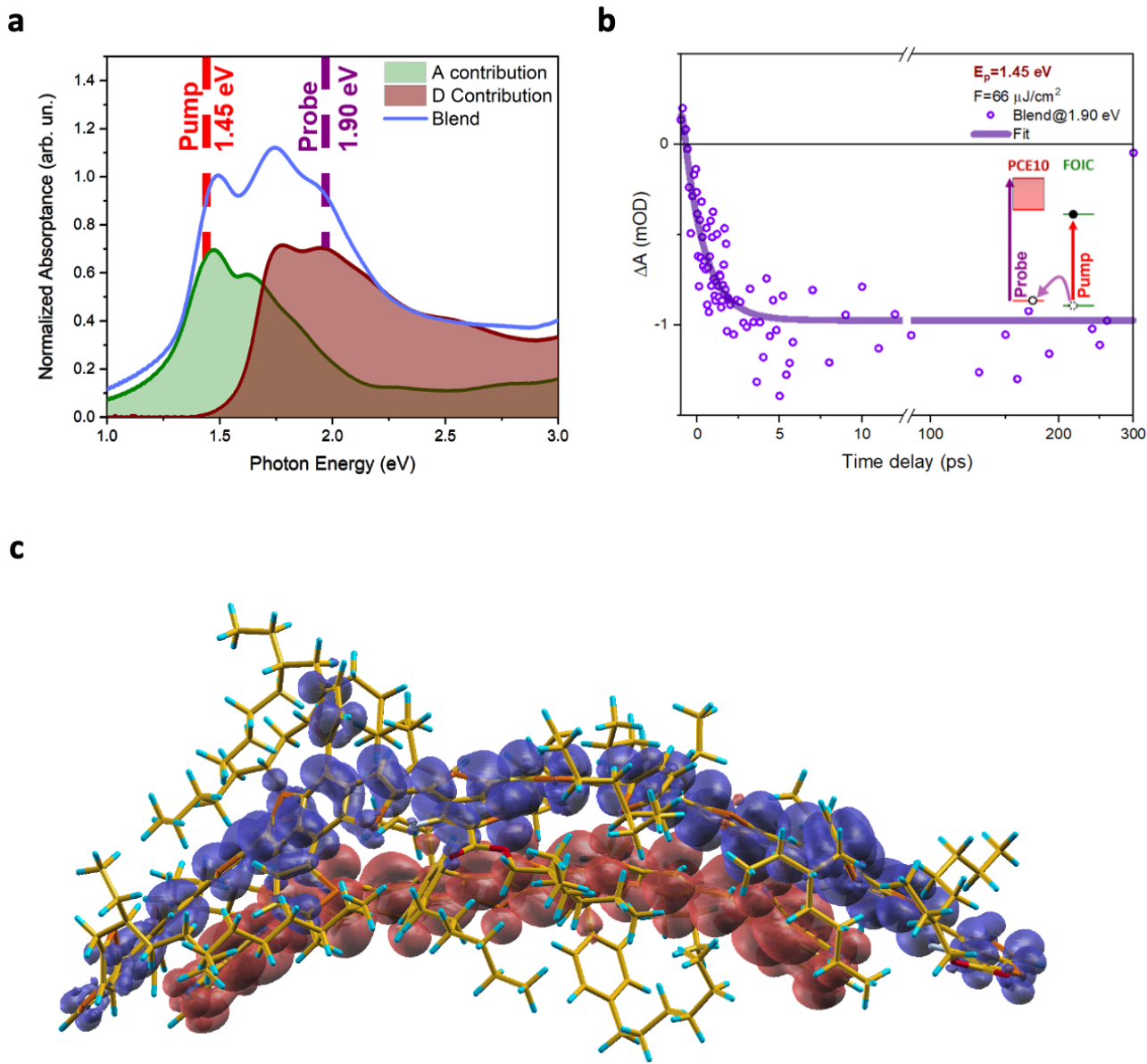

**Figure 5:** (a) Absorption in the Blend (blue line) and the A (green area) and the D (red zone) contribution at the absorption spectrum. (b) Temporal dynamics excited at the pump energy of 1.45 eV and the fluence of 66 µJ/cm² at the probe photon energy of 1.90 eV. Inset: simplified schematization of the hole separation process between D (PCE10) and A (FOIC). (c) Open shell calculation of the square modulus of the wavefunction of the explicitly introduced hole in the dyad FOIC:PCE10 (3), whereas the holes localized on the PCE10 (blue regions) and the electrons on FOIC (red regions).

The interpretation of FTAS results closely agrees with the theoretical result obtained on a FOIC/PCE10 heterojunction and shown in **Figure S1** (see the Experimental Section for calculations details). In the ground state of an interacting dyad formed by FOIC and a PCE10(4) tetramer, the HOMO is clearly localized on the polymer, while the LUMO is predominantly on the molecule (see the diagram of FOIC+PCE10(4) in **Figure S1**). When this system is excited, its lowest energy state S1 as described by TDDFT is a charge-separated state, with the electron localized on FOIC and the hole on the PCE10 tetramer. This is shown in **Figure 5c**, where the density difference between the ground state and the charge-separated state is represented, confirming the depletion of charge from PCE10 (blue regions) and the accumulation of charge on FOIC (red regions) upon excitation. The energy of this charge-separated state is 1.06 eV above the ground state, in agreement with the 1.03 eV offset between the valence and conduction bands estimated on the ground of measurements.

Based on the steady-state and TA results, the CS yield $\eta$ was estimated thanks to the following equation:

$$\eta = \frac{PB\,(1.90eV, 20ps)/A\,(1.90eV)}{PB\,(1.45eV, t_0)/A\,(1.45eV)} \qquad 2$$

where $PB(1.45eV, t_0)$ is the maximum value of the PB signal of the Blend at 1.45 eV probe energy, $PB(1.90eV, 20ps)$ is the intensity of the PB signal at 1.90 eV probe energy correlated to the separated charges and collected at 20 ps, and *A(eV)* is the Blend absorption at the selected probe photon energy. In this way, it was possible to estimate a charge separation efficiency of the blend of approximately 60 %, considering the intensity of PB signals normalized by the corresponding absorptions, all of which was considered proportional to the population of excited charges.

# Conclusions

Our study provides valuable insights into the optoelectronic properties and charge separation dynamics of a blend composed of FOIC and PCE10 organic semiconductor materials. By combining steady-state and time-resolved spectroscopic techniques with high-level DFT calculations, we characterize these materials as thin films both individually and in their blended state. In particular, the femtosecond transient absorption spectroscopy investigation revealed a reduction in the bimolecular exciton-exciton annihilation recombination rate in the acceptor when incorporated into the blend, compared to its pristine form. This reduction was attributed to a decreased exciton density within the acceptor, driven by the formation of an electron-hole charge separated state caused by the transfer of a hole from the acceptor to the donor polymer, as also confirmed by TDDFT simulations. This process was characterized by following the temporal evolution of the photobleaching signals associated with the excited states of the donor when the acceptor was selectively excited within the blend. In this way, the time constant of the hole transfer from the acceptor to the donor was quantified, yielding a value of (1.3±0.3) ps. Additionally, this study allowed the quantification of the exciton diffusion and the hole separation efficiency, offering valuable insights for the development of next-generation photovoltaic materials with optimized interfacial engineering strategies to further improve the charge separation and device efficiency.

# Acknowledgement

The authors acknowledge the European Project "Energy Harvesting in Cities with Transparent and Highly Efficient Window-Integrated Multi-Junction Solar Cells" (CITYSOLAR) for supporting the work, which received funding from the European Union's Horizon2020 research and innovation program under grant agreement number 101007084. G.A. and S.T. acknowledges the European Union – NextGenerationEU, M4C2, as part of the PNRR project NFFA-DI, CUP B53C22004310006, IR0000015. The work of G.M. has been financially supported by ICSC-Centro Nazionale di Ricerca in High Performance Computing, Big Data and Quantum Computing, funded by European Union-NextGenerationEU (grant CN00000013), and by the Italian Minister of the University and Research (MUR) within the PRIN-2022 research program (project "NIR+"). The authors acknowledge EUROFEL-ROADMAP ESFRI of the Italian Ministry of University and Research. The authors acknowledge Elettra Sincrotrone Trieste for providing access to its synchrotron radiation facilities and for financial support under the SUI internal project (proposal number 20220397).

# Supporting Information

## S1: Theoretical band diagram of FOIC, PCE10, and SC

On the basis of the theoretical results obtained inherent to the HOMO and LUMO levels, **Figure S1** shows the theoretical band diagram from FOIC, PCE10 (trimer, tetramer, and polymer obtained by periodic boundary conditions, labeled respectively as (3), (4), and (PBC)) and the FOIC+PCE10 (4) obtained with the B3LYP functional. The theoretical evaluation demonstrates that, concerning the energetics of frontier orbitals, even a tetramer is essentially converging with the PBC system, while a trimer is only partially satisfactory; for this reason, the calculation were performed on the FOIC:PCE10 (4) diad to ensure more reliable results.

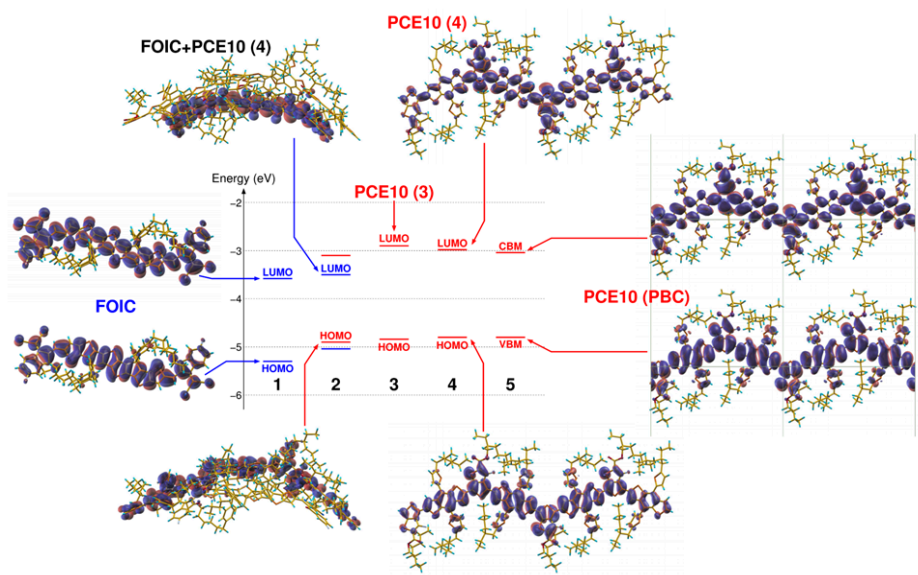

**Figure S1:** Theoretical band diagram for FOIC, PCE10 trimer (3), tetramer (4), and infinite crystal (PBC), and the PCE10(4): FOIC Blend.

Absorption spectra were also calculated on C09+VDWDF geometries in the sTDDFT framework[32] using the same B3LYP functional and the same basis sets for individual components and for a FOIC/PCE10 dyad in both trimer and tetramer cases. The simulations quite accurately replicate the spectroscopic features of the molecule, notably the strong near-infrared (NIR) absorption peaking at 869 nm (1.43 eV), while for PCE10 the pronounced absorption measured between red and NIR wavelengths shifts to lower energies with respect to the experimental absorption maximum (1.72 eV): the absorption peak is at 1.51 eV for the trimer and 1.39 eV for the tetramer. **Table S1** reports DFT calculation with B3LYP functional of the HOMO and LUMO energy levels and absorption peak energies. The theoretical data reported in the main text, in **Table S1** highlighted in blue, are FOIC, PCE10 (4) and FOIC+PCE10 (4) structures, as they result to be closer to the experimental results.

|  | HOMO (eV) | LUMO (eV) | Absorption peak (eV) |
|---|---|---|---|
| FOIC | -5.31 | -3.58 | 1.50 |
| PCE10 (PBC) | -4.80 | -3.06 |  |
| PCE10 (3) | -4.84 | -2.91 | 1.51 |
| PCE10 (4) | -4.80 | -2.98 | 1.39 |
| FOIC+PCE10 (3) | -4.89 | -3.48 | 1.61 |
| FOIC+PCE10 (4) | -4.89 | -3.50 | 1.46 |

**Table S1:** DFT calculation with B3LYP functional of the HOMO and LUMO energy levels and the theoretical absorption peak energies calculated on C09+VDWDF geometries in the sTDDFT framework.

## S2: Experimental photoelectron spectra and estimation of ionization energies of FOIC, PCE10, and Blend

The onset of the Valence Band (VB) of each material was defined as the intersection between the a linear fitting of the VB spectrum in the low binding energy region and a baseline, corresponding to the signal above the Fermi energy (dark counts) . Since the very broad experimental spectra do not allow to determine a unique onset energy, the straight line was fitted by varying the constrained axis limit between 0.7 and 3 eV with a fixed energy window of 500 meV. **Table S2** reports the mean value and the standard variation of binding energy onset of the VB, while **Figure S2** reports some selected fitting curves obtained with the previously described method.

On the basis of the experimental photoelectron spectroscopy (PES) and absorption spectra, the VB maximum and conduction band (CB) minimum was estimated as follows:

$$|VB| = |\Phi| + |E_b| \qquad \text{S1}$$
$$|CB| = |VB| - E_g$$

where Φ is the work function of the analyzer (4.51 eV), $E_b$ is the PES binding energy onset and $E_g$ is the optical bandgap onset.

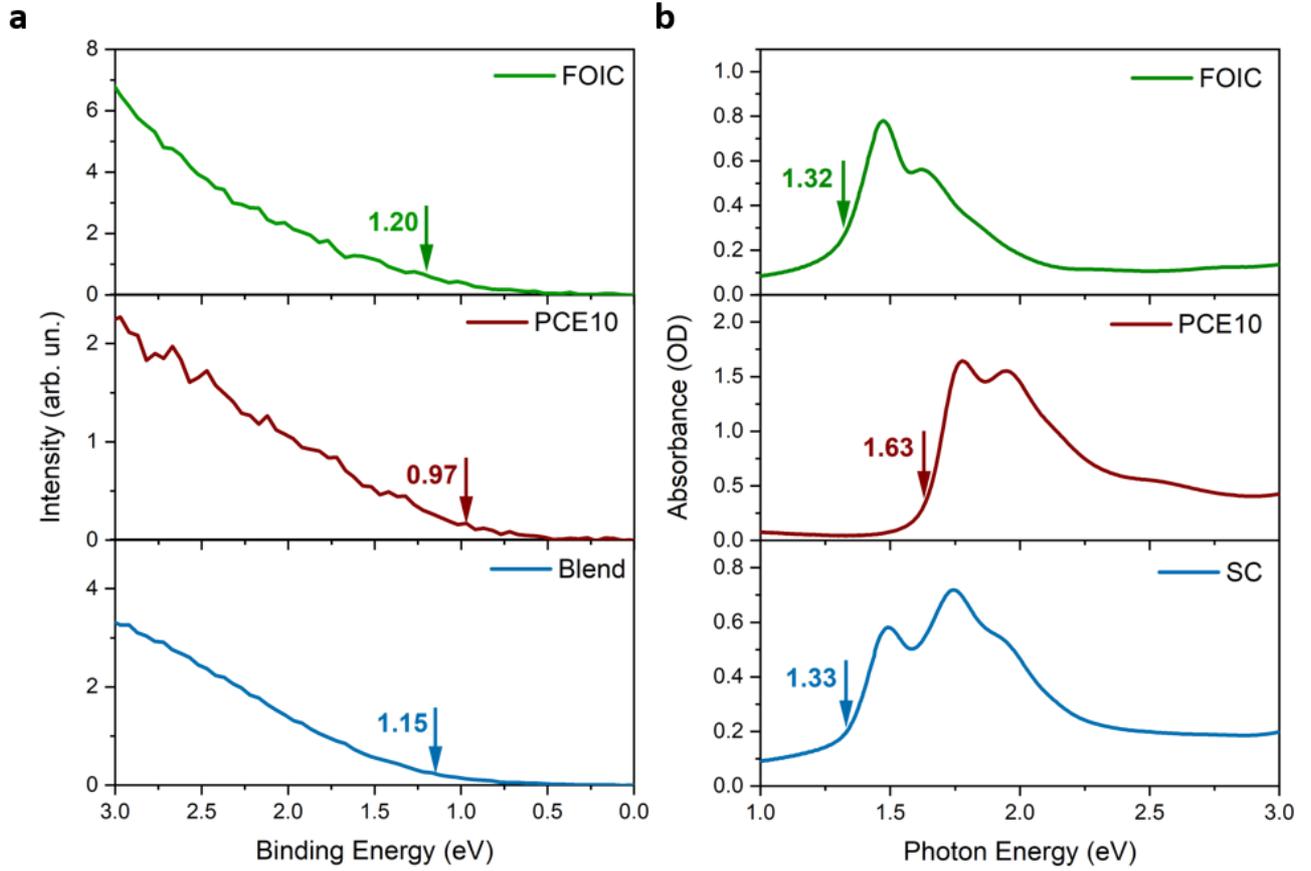

**Figure S2:** Experimental (a) PES spectra acquired at the photon energy of 75 eV and (b) absorption spectra obtained of the film of FOIC (green line), PCE10 (red line), and Blend (blue line).

All the values estimated using this method are reported in **Table S2**.

|            | FOIC  | Blend | PCE10 |
|------------|-------|-------|-------|
| $E_b$ (eV) | 1.20  | 1.15  | 0.97  |
| $E_{g}$ (eV) | 1.32  | 1.33  | 1.63  |
| VB (eV)    | -5.71 | -5.66 | -5.48 |
| CB (eV)    | -4.39 | -4.33 | -3.85 |

**Table S2:** Experimentally estimated $E_b$ and $E_g$, and the calculated VB maximum and CB minimum, for the FOIC, PCE10, and Blend.

## S3: Transient absorption spectroscopy

The exciton density generated by the pump excitation ($n_0$) is defined as follows:

$$n_0 = \frac{F}{\hbar\omega} \cdot \frac{A}{d}$$

Where F is the pump fluence, hω is the pump photon energy, A is the absorptance of the sample (i.e., the fraction of absorbed photons, see **Figure S3**), and d is the sample thickness. The thickness was estimated to be 50 nm for FOIC; 200 nm for PCE10; 105 nm for Blend. The values used for the estimation of the exciton density are reported in **Table S3**.

| Materials | Pump Energy (eV) | Absorptance (%) | Thickness (nm) |
|---|---|---|---|
| FOIC | 1.45 | 39 | 50 |
| Blend | 1.45 | 33 | 105 |
| PCE10 | 1.75 | 70 | 200 |

**Table S3**: Experimental values used for the estimation of the exciton density at the experimental pump photon energy and fluence.

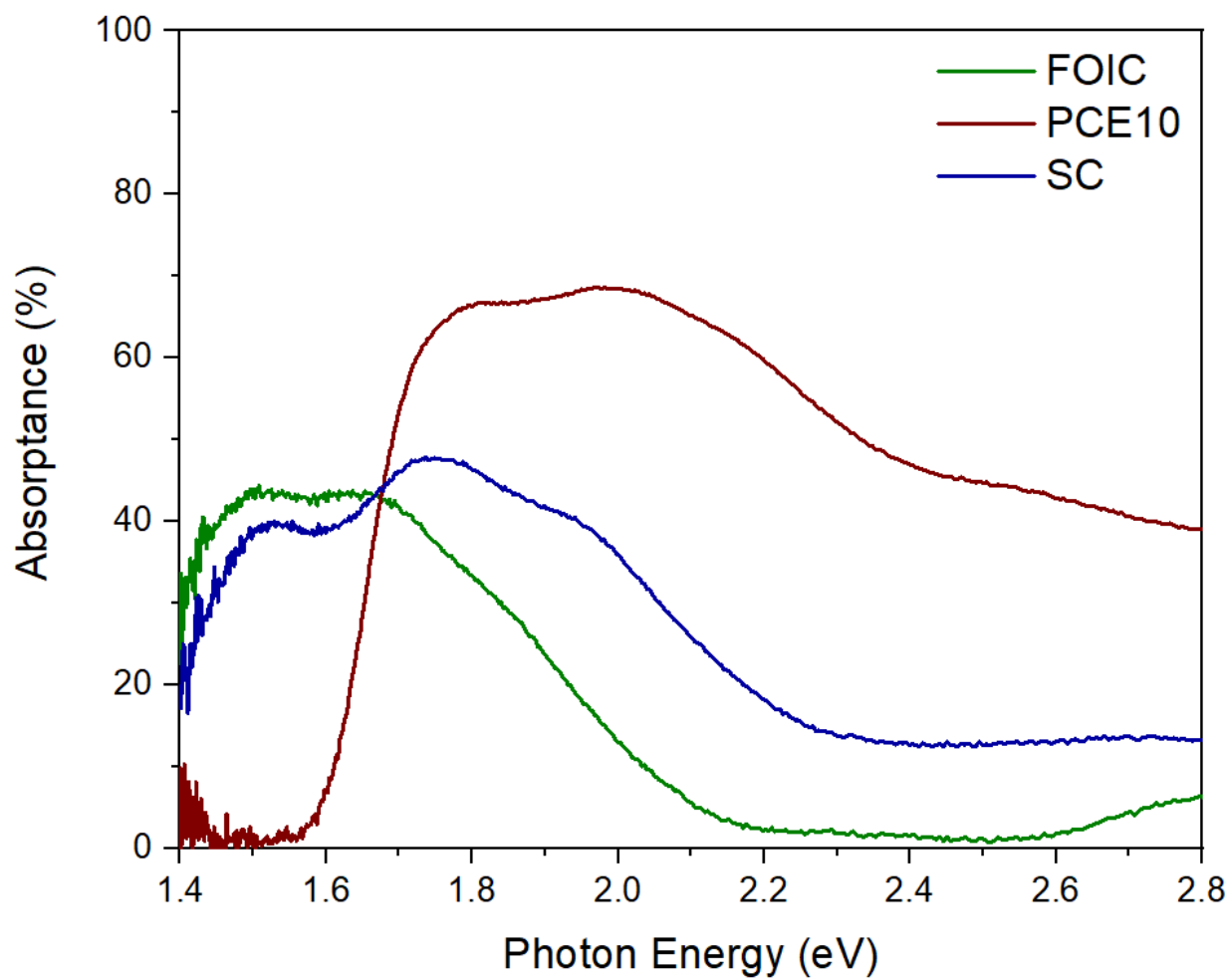

**Figure S3:** Absorptance spectra of FOIC (green line), PCE10 (red line), and Blend (blue line) obtained by the UV-2600 Shimadzu spectrometer with the multipurpose sample compartment add-on.

## S3.1 Fitting of the temporal dynamics

The fitting equations used to fit the exciton-exciton in this work is the following.

$$\Delta A(t) = \frac{\Delta A_2}{1 + \frac{t - t_0}{\tau}} + \Delta A_0 \quad \text{S2}$$

The bimolecular decay time τ extracted from the curve fitting is directly correlated to the bimolecular recombination rate γ by the following formula:

$$\frac{1}{\tau} = \gamma \cdot n_0 \quad \text{S3}$$

where $n_0$ is the exciton density.

The rise time of the PB signal was fitted with the following formula:

$$\Delta A_{rs}(t) = A_1 \cdot \left[ \left(1 + erf\left(\frac{(t-c)}{\sqrt{2}\sigma}\right)\right) - e^{\frac{\sigma^2}{2\tau_2^2} - \frac{(c+t)}{\tau_2}} \left(1 + erf\left(\frac{t-c}{\sqrt{2}\sigma} - \frac{\sigma}{\sqrt{2}\tau_2}\right)\right) \right] + A_o \quad \text{S4}$$

where $\Delta A_{rs}(t)$ is the rising temporal trend of the transient signal collected at selected probe energy, $A_1$ is the amplitude of the transient signal, $c$ is the time where the maximum of the laser pulse takes place, the temporal form of the pulse is described by a Gaussian with a standard deviation of $\sigma \approx 0.03 ps$, t is the time delay between pump and probe, $\tau_2$ is the rise time of the transient signal and $A_o$ is the offset which describes the mean transient intensity value at negative times.

## S3.2 FTAS results at the pump energy of 1.72 eV

**Figure S4a** shows the measured TA spectra obtained with a pump photon energy of 1.72 eV and at a fluence of 220 µJ/cm² for FOIC, PCE10, and Blend. The TA spectra show PB signals where the absorption spectra show intense peaks. In fact, FOIC shows an intense PB signal at 1.45 eV and PCE10 at 1.72 eV. Both systems show also negative features at higher energies that were attributed to the vibrational state progressions already discussed in the main text. On the other hand, the Blend exhibits feature that can be readily identified with those found in the TA spectra of the individual components: a strong PB signal at 1.72 eV, associated with a donor-like transition; a less intense PB signal at 1.45 eV, associated with an acceptor-like transition. The different intensity of the PB signals since the energy of the pump (1.72 eV) is almost resonant with the donor-like optical transition, inducing a more substantial change in the electronic occupation of the states involved in the excitation.

**Figure S4b** shows the temporal dynamics obtained at the pump photon energy of 1.75 eV with a fluence of 220 µJ/cm² and at the probe photon energy of 1.72 eV for PCE10 and of 1.45 eV for FOIC, namely at the energies of the PB signal minima.

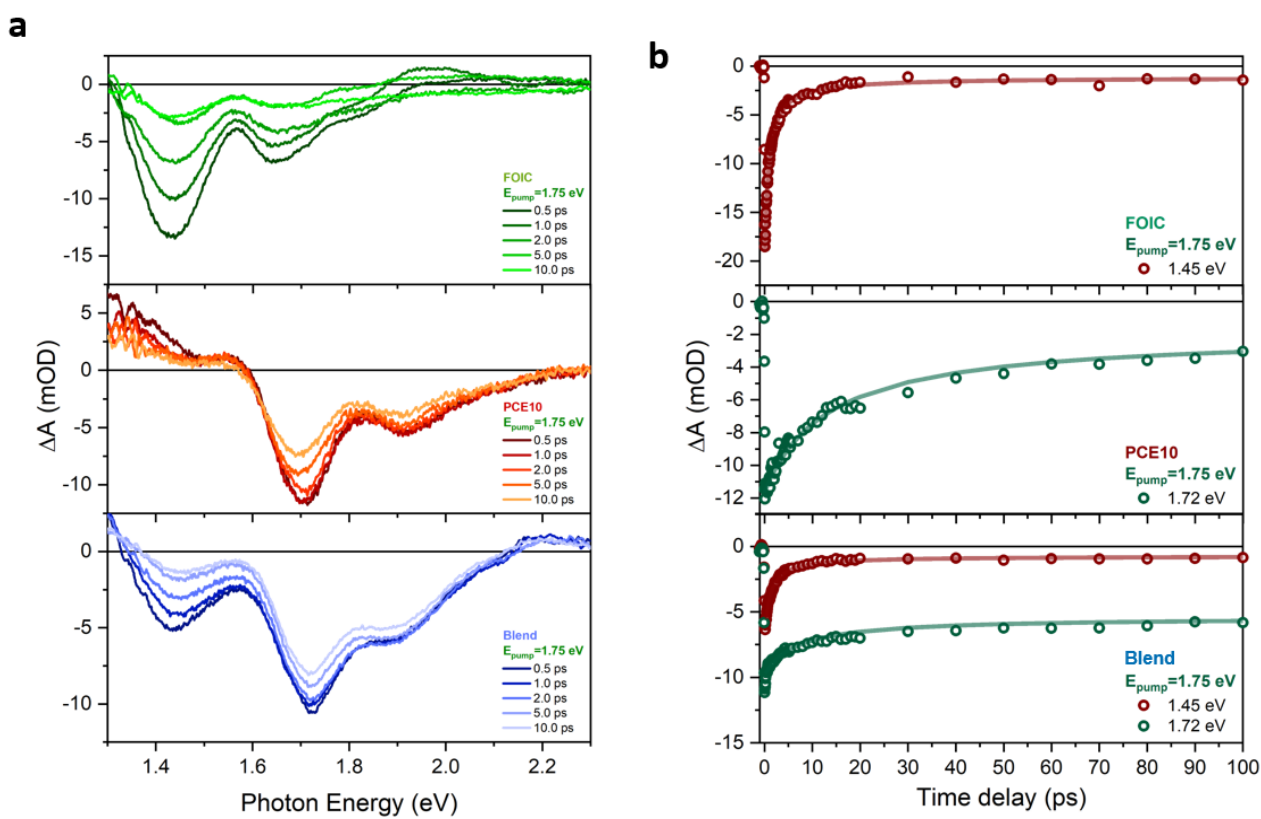

**Figure S4:** (a) TA spectra were obtained at the pump fluence of 220 µJ/cm2 and the pump photon energy of 1.75 eV for PCE10 (red), FOIC (green) and Blend (blue); (b) temporal dynamics obtained at the pump fluence of 220 µJ/cm² and the pump photon energy of 1.75 eV for PCE10 (green scatter the temporal dynamics at the probe photon energy of 1.72 eV), FOIC (red scatter the cuts at the probe photon energy of 1.45 eV), and Blend (red and green scatter the cuts at the probe photon energy of 1.45 eV and 1.72 eV respectively).

The Blend was studied in the same excitation conditions and the temporal dynamics were explored at the probe energies selected for PCE10 and FOIC. The experimental results demonstrates that the decay of the PB signals lasts for tens of picoseconds, in agreement with the observed trend in other organic thin films[18–20] and their decay was attributed to EEA process. The estimated time decay for FOIC, PCE10, and Blend are reported in **Table S4.**

|  | FOIC | Blend | PCE10 | Blend |
|---|---|---|---|---|
| *Probe photon energy (eV)* | 1.45 | 1.45 | 1.72 | 1.72 |
| *τ(ps)* | 0.92±0.05 | 1.12±0.07 | 13.5±1.5 | 6.9±1.1 |

**Table S4**: EEA decay time obtained for FOIC, PCE10, and Blend at the pump photon energy of 1.75 eV obtained at the different probe photon energy.

The decay times obtained by the fit procedure show that FOIC exhibits a faster decay with respect to PCE10. Additionally, the Blend shows similar trends but not identical to those recorded from the pristine FOIC and PCE10. This behavior is ascribable to the fact that under this experimental condition both the donor and acceptor are excited in the Blend, making it difficult to disentangle the effect of the excitation of the single component.

To gain a clearer understanding of the recombination mechanisms underlying the different temporal dynamics, FTAS measurements were performed on FOIC and Blend at a pump photon energy of 1.45 eV. In this way, only the acceptor in the Blend was excited, giving the opportunity to follow the charge dynamics of a single component of the Blend (see the main t